\documentclass[aps]{revtex4}
\usepackage{amssymb}
\usepackage[dvips]{graphicx}
\usepackage{amsmath}
\usepackage{graphicx}
\usepackage{makeidx}
\usepackage{subfigure}

\begin{document}

\title{Melting Transition of Vortex Lattice in Point Vortex Systems}
\author{Hidetsugu Sakaguchi and Kenji Kita}
\address{Department of Applied Science for Electronics and Materials,\\
Interdisciplinary Graduate School of Engineering Sciences,\\
Kyushu University, Kasuga, Fukuoka 816-8580, Japan}

\begin{abstract}
Point vortices take a triangular lattice structure in a rotating system as a minimum energy state. We perform a numerical simulation of point vortex systems using initial conditions indicating that the triangular lattice is randomly perturbed.  
The total energy increases with the magnitude of the perturbation. 
When the energy is increased, the vortex lattice becomes irregular and a layered structure appears. When the energy is further increased, the layered structure disappears and a liquidlike state appears. 
We interpret the melting transition with a mean-field approximation for layered structures.   
\end{abstract}

\maketitle
\section{Introduction}
Quantized vortices are important in several quantum systems such as superfluid $^4$He and superconductors. Triangular lattices of vortices and magnetic fluxes were observed in rotating superfluids and type II superconductors in magnetic fields. Similar quantized vortex lattices have recently been found  in rapidly rotating Bose-Einstein condensates of ultracold diluted trapped gases.\cite{rf:1,rf:2} The quantized vortex states in the Bose-Einstein condensates have been intensively studied theoretically.\cite{rf:3,rf:4,rf:5} 
The vortex lattice state is considered to be the lowest energy state in the rotating Bose-Einstein condensates with repulsive interaction. The Gross-Pitaevskii equation describes fairly well the Bose-Einstein condensates.  The Gross-Pitaevskii equation is an equation in which the total energy and the angular momentum are conserved. If the initial conditions deviate from the vortex lattice state, various types of complex dynamics are expected to occur. We studied the time evolution of the Gross-Pitaevskii equation from an initial condition analogous to a rigid rotation, and found that a vortex lattice state appeared owing to  Kelvin-Helmholtz instability.\cite{rf:6}  Since the obtained state is not the lowest-energy state, the vortices exhibited an irregular rotating motion around the regular lattice points.  When the irregular motion becomes large, the lattice structure is expected to melt into a liquid state. 

Several phenomena similar to the melting phenomena of vortex lattices have been studied in other experiments. A liquid-to-crystal transition was studied in a two-dimensional sheet of electrons~\cite{rf:7} and a transition from turbulence to vortex crystals was studied in magnetized electron systems.\cite{rf:8} The lattice structure and its melting transition in charged-particle systems have been intensively studied.\cite{rf:9,rf:10} 
The melting transitions from vortex lattices to vortex liquids have also been studied in the type II superconductors.\cite{rf:11,rf:12,rf:13} Vortex-lattice melting by quantum fluctuations in Bose-Einstein condensates on a one-dimensional optical lattice was theoretically studied.\cite{rf:14} The Lindemann criterion of lattice vibration is often used whether the lattice is melted or not in many studies.

On the other hand, point vortex systems have been intensively studied in fluid mechanics, especially as an ideal system of two-dimensional turbulence. 
The integrable, chaotic, and turbulent motions were found in some point vortex systems.\cite{rf:15} The statistical mechanics of point vortices have been studied since the pioneering work by Onsager.\cite{rf:16,rf:17,rf:18}  

In Bose-Einstein condensates, the vortex system can be approximated by a point vortex system if the repulsive interaction is sufficiently strong and the effect of the vortex core is negligible. In this paper, we discuss a melting transition of vortex lattices in general point vortex systems from the viewpoints of nonlinear dynamics and statistical mechanics.  

We perform some direct numerical simulations of point vortex systems and find that there is a transition from a layered distribution to a uniform distribution for the positional distribution of point vortices.  We interpret the transition as the melting transition and study the transition using a mean-field approximation for the layered distribution. A similar layered structure and the melting transition were found in a two-dimensional system of charged particles using Monte-Carlo simulations;\cite{rf:9} however, a melting transition in point vortex systems has not been numerically studied.  The melting transition in a system of charged particles is somewhat similar to that in our model, because layered structures appear and the layered structures are destroyed at melting transitions. However, the dynamics of our point-vortex system is different from that in charged particles, in that the conservation law of angular momentum plays an important role in the point-vortex system.  The equilibrium distribution is different from a simple canonical distribution owing to the additional conservation law. For example, an average rotational flow can exist even in an equilibrium state in the point vortex system, although such average flow is impossible in typical equilibrium states. 
We consider that the point vortex system is an interesting model of statistical mechanics in that long-range interactions and the additional conservation law are essential.

\section{Models of Point Vortex Systems} 
We study point vortex systems because the equation of motion is simple, and a large-scale numerical simulation is possible in contrast to the numerical simulation of partial differential equations such as the Navier-Stokes and Gross-Pitaevskii equations. The point vortex systems are suitable for quantized vortices in superfluids because the circulation is quantized to a definite constant value, while the vorticity field changes continuously in most classical fluids. 
In three dimensions, a point vortex becomes a vortex filament. Vortex filaments were used to study  vortex motions in superfluids and superfluid turbulences theoretically.\cite{rf:19} It is known that the approximation by point vortices and vortex filaments of vortices in superfluids is fairly good. 
Point vortices in two dimensions obey the equation of motion as explained in textbooks of fluid mechanics~\cite{rf:20}: 
\begin{eqnarray}
\frac{dx_i}{dt}&=&-\frac{1}{2\pi}\sum_{j\ne i}\frac{\kappa_j(y_i-y_j)}{(x_i-x_j)^2+(y_i-y_j)^2},\nonumber\\
\frac{dy_i}{dt}&=&\frac{1}{2\pi}\sum_{j\ne i}\frac{\kappa_j(x_i-x_j)}{(x_i-x_j)^2+(y_i-y_j)^2},
\end{eqnarray}
where $i=1,2,\cdots, N$, $N$ is the total number of vortices, $(x_i,y_i)$ denotes the two-dimensional position of the $i$th vortex, and $\kappa_i$ is the circulation of the $i$th vortex. Point vortices interact with other vortices through a long-range force equivalent to the two-dimensional Coulomb force; however, the equation of motion is not Newton's equation of motion as in a system of charged particles. Each point vortex just flows passively under the velocity field generated by other point vortices. 
For quantized vortices in Bose-Einstein condensates, the circulation $\kappa_i$ takes discrete values $\kappa_i=2\pi n\hbar/m$, where $n$ is an integer and $m$ denotes the atomic mass. In most cases, $n$ takes $+1$ or $-1$, because a vortex of large $n$ tends to break up into $n$ vortices of circulation $2\pi\hbar/m$.  If the circulation $\kappa_i$'s are all positive, an anticlockwise rotational flow of average angular velocity $\Omega$ is induced on the average. 
The average rotational flow seems to disappear, if the system is observed from a rotational frame of angular velocity of $-\Omega$.  In the rotational frame, the equation of motion is expressed as 
\begin{eqnarray}
\frac{dx_i}{dt}&=&\Omega y_i-\frac{1}{2\pi}\sum_{j\ne i}\frac{\kappa (y_i-y_j)}{(x_i-x_j)^2+(y_i-y_j)^2},\nonumber\\
\frac{dy_i}{dt}&=&-\Omega x_i+\frac{1}{2\pi}\sum_{j\ne i}\frac{\kappa (x_i-x_j)}{(x_i-x_j)^2+(y_i-y_j)^2}.
\end{eqnarray}
Here, we have assumed that all the point vortices have the same circulation $\kappa_i=\kappa=2\pi \hbar/m$. 
If the spatial scale transformation $x^{\prime}=x/R,y^{\prime}=y/R$, where $R$ is the length scale such as the radius of a container used to confine the vortices, and the temporal scale transformation $t^{\prime}=\kappa t/(2\pi R^2)$ are assumed, 
the equation of motion of point vortices is written in a dimensionless form as  \begin{eqnarray}
\frac{dx_i}{dt}&=&\omega y_i-\sum_{j\ne i}\frac{y_i-y_j}{(x_i-x_j)^2+(y_i-y_j)^2},\nonumber\\
\frac{dy_i}{dt}&=&-\omega x_i+\sum_{j\ne i}\frac{x_i-x_j}{(x_i-x_j)^2+(y_i-y_j)^2},
\end{eqnarray}
where $x_i^{\prime},y_i^{\prime}$, and $t^{\prime}$ are rewritten respectively as $x_i,y_i$ and $t$, and $\omega=2\pi R^2\Omega/\kappa$. The energy (Hamiltonian) $E$ in this system is expressed as
\begin{equation}
E=-\frac{1}{2}\sum_{i}\sum_{j\ne i}\ln r_{i,j}+\sum_i\frac{1}{2}\omega r_i^2,
\end{equation}
where $r_i=\sqrt{x_i^2+y_i^2}$ and $r_{i,j}=\sqrt{(x_i-x_j)^2+(y_i-y_j)^2}$. 
(The energy is expressed as $\hbar^2/(m R^2) E$ if dimensions are recovered.)
Equation (3) can be expressed using $E$ as 
\begin{equation}
\frac{dx_i}{dt}=\frac{\partial E}{\partial y_i}, \;\frac{dy_i}{dt}=-\frac{\partial E}{\partial x_i}.
\end{equation}
In this time evolution, the total energy $E$ is conserved. 
The angular momentum $L=\sum_i\{x_i(dy_i/dt)-y_i(dx_i/dt)\}$ is also conserved in the time evolution of eq.~(3).
 The center of mass  $X=(1/N)\sum_ix_i$ and $Y=(1/N)\sum_iy_i$ obeys $dX/dt=\omega Y$ and $dY/dt=-\omega X$. 

In the approximation of point vortices, the effect of the vortex core is neglected. If the core size is sufficiently smaller than the average interval between neighboring vortices, the approximation might be good. 
The effect of the confinement by a harmonic potential as often used in Bose-Einstein condensates is not easily incorporated in point vortex systems. 
However, the confinement by a hard wall potential at $r=R$ can be incorporated in  point vortex systems by using antivortices at the mirror-image positions $R^2/r_i$, but for simplicity, we do not consider the effect in this paper.

The point vortices including $N$ vortices in a rotating container take a triangular lattice as a stable configuration.  The energy $E$ is minimized at the stable configuration. The stable configuration can be numerically obtained using the equations:
\begin{eqnarray}
\frac{dx_i}{dt}&=&-\omega x_i+\sum_{j\ne i}\frac{x_i-x_j}{(x_i-x_j)^2+(y_i-y_j)^2},\nonumber\\
\frac{dy_i}{dt}&=&-\omega y_i+\sum_{j\ne i}\frac{y_i-y_j}{(x_i-x_j)^2+(y_i-y_j)^2}.
\end{eqnarray}
This is because eq.~(6) is expressed as $dx_i/dt=-\partial E/\partial x_i$ and $dy_i/dt=-\partial E/\partial y_i$, and the energy $E$ decreases with time, and a state of the lowest energy is obtained as a stationary state in the time evolution of eq.~(6). 
The position of the $i$th vortex at the lowest energy is expressed as $(x_{i0},y_{i0})$.  Vortices are attracted to the origin $(0,0)$ by the first term in eq.~(6) and they interact with each other via the repulsive force expressed by the second term in eq.~(6). As a result, the vortices find a stationary configuration such as a triangular lattice. 
Campbell and Ziff found various stable configurations of $N$ point vortices.\cite{rf:21,rf:22}. Stationary solutions to eq.~(6) are stationary solutions to eq.~(3). Finding stationary solutions to eq.~(3) through eq~(6) is a useful method.  

The angular momentum $L=\sum_i\{x_i(dy_i/dt)-y_i(dx_i/dt)\}$ can be calculated as \[L=-\sum_i\{\omega (x_i^2+y_i^2)\}+\sum_{i,j}\frac {x_i(x_i-x_j)+y_i(y_i-y_j)}{(x_i-x_j)^2+(y_i-y_j)^2}=-\omega I+\frac{N(N-1)}{2},\] where $I=\sum_i r_i^2$. At the stationary states, the quantity $I$ satisfies $I=N(N-1)/(2\omega)$, because $L=0$. Assume that the point vortices take a triagular lattice at the stable configuration. If the interval between the nearest-neighbor vortices is denoted as $d$, the area of the elemental triangle of the lattice is expressed as $S=\sqrt{3}d^2/4$. 
If the average radius of the entire triangular lattice is denoted as $R_0$ and the average density of vortices is denoted as $\rho$,  $\rho$ is expressed as $(3/6)/S=2\sqrt{3}/(3d^2)$, the total vortex number is $N=\pi R_0^2\rho$, and $I$ is expressed as $I=N(N-1)/(2\omega)=\pi R_0^4\rho/2$. From these relations, typical length scales are evaluated as $R_0=\sqrt{(N-1)/\omega}$ and $d=(2\pi/\omega)^{1/2}(1/3)^{1/4}$.  

If the point vortices are mixed up well by the high-dimensional chaos and there is no spatial correlation, we can apply the Debye-H\"uckel theory of electrolyte solution \cite{rf:22} for the distribution function $P(r)$, although  there might be a criticism that a canonical distribution could not be applied to a small system of vortex number $N=7$.   In the Debye-H\"uckel theory, the density $P(r)$  depends only on the radius $r$ from the center. The density of the electrolyte solution and the electric potential are determined self-consistently. That is, the electric potential is determined by the density of the electrolyte solution through the Poisson equation, and the density is determined by the canonical distribution under the electric potential. The density $P(r)$ of the electrolyte solution around an electrode obeys the relation  $P(r)\propto \exp\{-\beta \phi(r)\}$, where $r$ is the distance from the electrode, $\phi$ is the electric potential, $\beta=q/(k_BT)$, $q$ is the charge of ions, $k_B$ is the Boltzmann constant, and $T$ is the temperature.  Similarly, the stationary distribution $P(r)$ of point vortices is expected to obey a canonical distribution,
\begin{equation}
P(r)=\exp\{-\beta \phi(r)-\gamma (1/2)\omega r^2\}/Z,
\end{equation}
where $\phi(r)$ is an effective potential for the point vortex, and $Z$ is determined from the normalization condition $\int_0^{\infty}P(r)2\pi rdr=N$. 
The two parameters $\beta$ and $\gamma$ are determined from the conservation laws of the interaction energy $E_1=-\frac{1}{2}\sum_{j\ne i}\ln r_{i,j}$ and the rotational energy $E_2=(1/2)\omega\sum r_i^2$. If $\beta=\gamma$, $P(r)$ has the form of the standard canonical distribution $P(r)=\exp(-\beta E)/Z$ where $E=E_1+E_2$. However, $\gamma$ takes a different value from $\beta$ in our system. 
The distribution is a generalization of the distribution in the Debye-H\"uckel theory in that the conservation law of angular momentum is taken into consideration. In our system, the interaction energy is specified and the effective inverse temperature $\beta$ is determined from the conservation law of $E_1$. The two quantities 
$E_1$ and $E_2$ are expressed in the mean-field approximation as
\begin{eqnarray}
E_1&=&-\int_0^{\infty}P(r)Q(r)(\ln r)2\pi rdr,\nonumber\\
E_2&=&(1/2)\omega\int_0^{\infty}r^2P(r)2\pi r dr.
\end{eqnarray}   
Here, we have assumed that $P(r)$ is equal to the density of the point vortex, and $Q(r)$ is defined as $Q(r)=\int_0^{r}P(r^{\prime})2\pi r^{\prime}dr^{\prime}$. That is, $E_1$ expresses the integration of the interaction energy of the point vortex at $r$ with all the point vortices within a circular region of radius $r$. $E_2$ expresses the rotational energy using the vortex density $P(r)$.
The effective potential $\phi$ satisfies 
\begin{equation}
\nabla^2\phi=\frac{1}{r}\frac{\partial}{\partial r}\left (r\frac{\partial\phi}{\partial r}\right )=-2\pi P.
\end{equation}
This equation is equivalent to the Poisson equation in two dimensions.   
It is because the logarithmic interaction among point vortices is equivalent to that among two-dimensional point charges, and the electric potential obeys the Poisson equation under the distribution of point charges.  
The quantity $E_2$ is fixed to be $N(N-1)/4$, because $I=\sum_ir_i^2$ is equal to $I=N(N-1)/(2\omega)$ owing to the conservation law of the total angular momentum. 
The parameter $\gamma$ is determined as $E_2$ is equal to $N(N-1)/4$.  
The above equations can be numerically solved. If $\beta=\infty$ or the temperature is zero, $P(r)$ takes the form of a step function: $P(r)=\omega N/\{\pi(N-1)\}$ for $r<r_0=\sqrt{(N-1)/\omega}$, and $P(r)=0$ for $r>r_0$. 
The length $r_0$ is the same as $R_0$ previously evaluated for the triangular lattice in the lowest energy. 
This type of mean-field approximation was studied previously and reported in refs.17 and 18. 

We will study the dynamical and statistical behaviors of point vortex systems using eq.~(3). Although the energy (Hamiltonian) (4) seems to depend on $r$ explicitly, the system is effectively uniform, because the term $(1/2)\omega r^2$ is cancelled by the repulsive interaction among vortices.  Assuming the uniformity, we use initial conditions randomly perturbed from the regular triangular lattice. That is,  vortices are  perturbed from the stable configuration $(x_{i0},y_{i0})$  as  $(x_{i}(0),y_{i}(0))=(x_{i0}+r_{xi},y_{i0}+r_{yi})$, where $r_{xi}$ and $r_{yi}$ are random numbers randomly chosen from  a uniform probability distribution between $-r_0$ and $r_0$. The standard deviation $\Delta r_0=\{\sum (r_{xi}^2+r_{yi}^2)/N\}^{1/2}$ is evaluated as $\Delta r_0=\sqrt{2/3}r_0$. 
The parameter $r_0$ denotes the magnitude of the random perturbation. 
We have further adjusted  the random numbers as $\sum_i r_{xi}=\sum_i r_{yi}=0$ and $\sum_i\{(x_{i0}+r_{xi})^2+(y_{i0}+r_{yi})^2\}=N(N-1)/(2\omega)$. 
Under these initial conditions, the center of mass is always $(0,0)$ and the angular momentum is kept to be $L=0$.  The total energy $E$ increases with $r_0$.  We have used these random initial conditions to change the total energy. 
 As $E$ is increased, chaotic dynamics is observed. The ergodicity and a thermal equilibrium state are expected, as in the molecular dynamics (MD) simulation using Newton's equation of motion. 
Our numerical simulation corresponds to the MD simulation of the lattice vibration of a two-dimensional triangular lattice where initial conditions are set to be in a randomly perturbed lattice structure.  
The temperature of the point vortex system increases with the internal energy $E$; however, no explicit form of the temperature is known in the point vortex system, while the temperature is expressed using the temporal average of kinetic energy in the MD simulation. We have performed numerical simulations of systems including various total numbers of vortices. We show a typical example of $N=37$ in this paper to show the melting transitions from the viewpoint of dynamical systems including those with a relatively small total number of vortices.    
\begin{figure}[t]
\begin{center}
\includegraphics[height=4.cm]{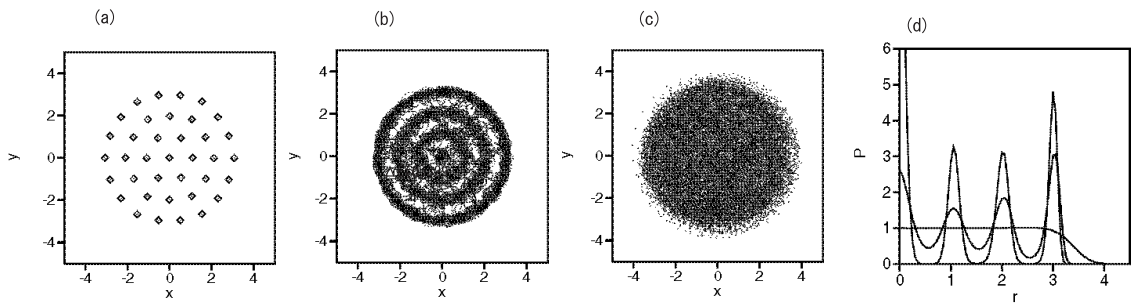}
\end{center}
\caption{(a) Vortex configuration at the lowest energy state at $E=E_0=-370.33$ in a point-vortex system of 37 vortices. Plots of $(x_i(t),y_i(t))$ at $t=0.2\times n$ ($n=1,2,\cdots, 750$) at (b) $E=-369.522$, and  (c) $E=-361.055$. (d) Distribution $P(r)$ at $\Delta E=0.329,0.809$ and 9.27.
}
\label{f1}
\end{figure}
\section{Thirty-Seven Vortex System}
In this section, we study a point vortex system with thirty-seven vortices at $\omega=3$. Figure 1(a) displays a vortex configuration at the lowest energy state at $E=E_0=-370.33$.
Figures 1(b) and 1(c) show plots of $(x_i(t),y_i(t))$ at $t=0.2\times n$ ($n=1,2,\cdots, 750$) at (b) $E=-369.522$, and  (c) $E=-361.055$.
 At $E=-369.522$, a four-layered structure appears. At $E=-361.055$, the layered structure is broken and a uniform distribution appears. 
Vortices come and go between the outer three layers for $\Delta E=E-E_0>0.301$. 
Figure 1(d) shows the distribution $P(r)$ of the vortices in the radial direction at $\Delta E=0.329,0.809$ and 9.27 by the numerical simulation. The peak structures of $P(r)$ at $\Delta E=0.329$ and 0.809 correspond to the layered structures, and the flat distribution at $\Delta E=9.27$ corresponds to the liquidlike state. 

\begin{figure}[t]
\begin{center}
\includegraphics[height=4.cm]{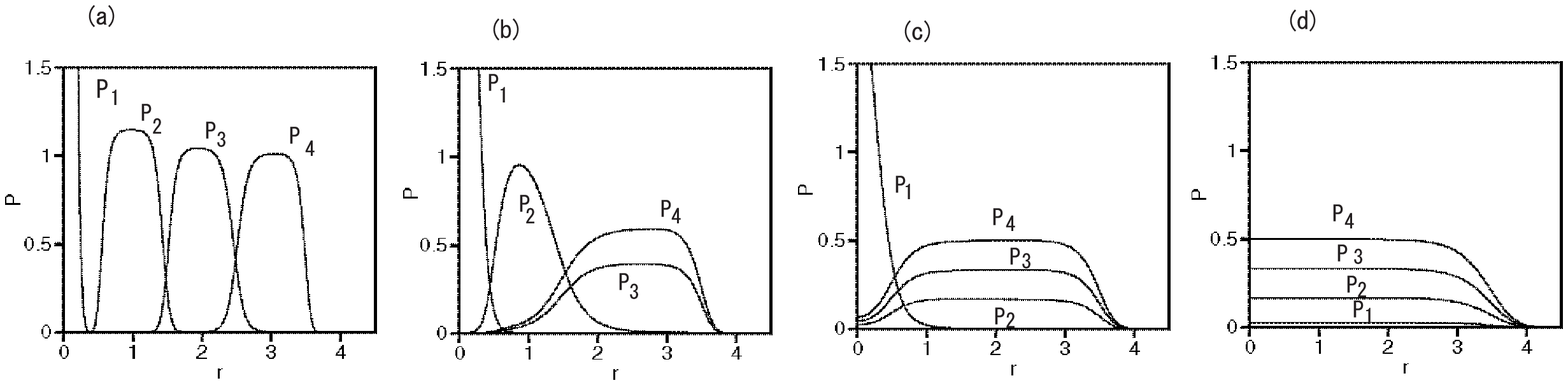}
\end{center}
\caption{Stationary solutions of the mean-field equation (10) at  (a) $\beta=42.5$, (b) $\beta=7.5$, (c) $\beta=5$, and (d) $\beta=2.5$.
}
\label{f2}
\end{figure}

The melting transition can be treated qualitatively by the mean-field approximation even for this four-layered structure. The probability distributions for the four layers are expressed as $P_1,P_2,P_2$, and $P_4$, and the potentials are denoted as $\phi_1,\phi_2,\phi_3$, and $\phi_4$. 
\begin{eqnarray}
P_i(r)&=&\exp\{-\beta \phi_i(r)-\gamma (1/2)\omega r^2\}/Z_i,\nonumber\\
E_1&=&-\int_0^{\infty}\{P(r)Q(r)-\sum_{i=1}^4 q_iP_i(r)Q_i(r)\}(\ln r)2\pi rdr,\nonumber\\
E_2&=&(1/2)\omega\int_0^{\infty}r^2P(r)2\pi r dr,\nonumber\\
\nabla^2\phi_i&=&\frac{1}{r}\frac{\partial}{\partial r}\left (r\frac{\partial\phi_i}{\partial r}\right )=-2\pi (P-q_iP_i),
\end{eqnarray} 
where $Z_i$'s are normalization constants for $P_i$, $Q_i=\int_0^rP_i(r^{\prime})2\pi r^{\prime}dr^{\prime}$, $P(r)=\sum_{i=1}^4P_i(r)$, and $Q(r)=\sum_{i=1}^4Q_i(r)$. The quantities $q_i$'s denote the proportion of self-interaction in each layer: $q_1=1,q_2=1/6,q_3=1/12$, and $q_4=1/18$. Stationary solutions to eq.~(10) were numerically obtained.    
 Figure 2(a) shows the four-layered solution for $\beta=42.5>\beta_1$. 
The interval between the layers is about 0.95, which is comparable to the result of the direct numerical simulation. The layered structure disappears from the outer layers. 
The outer two layers are merged into one layer in Fig.~2(b) at $\beta=10.3<\beta_1$. 
The outer three layers are merged into one layer in Fig.~2(c) at $\beta=5.9<\beta_2$. The layered structure disappears completely and changes into a liquidlike state  at $\beta=2.5<\beta_3$. 

\section{Summary}
We have performed a direct numerical simulation of point vortex systems.
We have found that layered structures appear when the energy increases. 
The layered structures are gradually broken and a uniform distribution for the position of vortices is realized when the energy increases sufficiently. 
Taking the layered structure into consideration, we have derived a mean-field equation for the distribution $P_i$ for each layer. The distribution is assumed to obey a canonical distribution based on the two conservation laws of the total energy and the total angular momentum.  The interval between the neighboring layers can be evaluated in the mean-field theory for the layered structure.  The melting transition from a layered structure to a liquid state appears naturally in the mean-field approximation.

\end{document}